\documentclass{acm_proc_article-sp}

\usepackage{graphicx}
\usepackage{colordvi}
\usepackage{url}
\usepackage{listings}
\usepackage{textcomp}
\usepackage{algorithm}
\usepackage{amsmath}
\usepackage{algorithm}
\usepackage{algpseudocode}
\usepackage[english]{babel}
\usepackage{epstopdf}
\usepackage[T1]{fontenc}
\usepackage{xcolor}

\begin{document}

\title{Towards Large-Scale Autonomous Wireless Sensor Networks}

%
%
%
%
%
\numberofauthors{1} 
%

\author{
\alignauthor
Francesco Fraternali\\
       \affaddr{University of California, San Diego}\\
       \email{\footnotesize frfrater@eng.ucsd.edu}
}

\maketitle
\begin{abstract}
Wireless Sensor Networks (WSNs) have the goal of gathering data from the environment. The advent of the Internet of Things (IoT) drastically changed WSN's vision that, as never before, needs to expand and include hundreds or thousands of sensors. But to follow the current IoT trends new techniques need to be implemented since orders of thousands of sensor nodes are not manageable by today's WSNs systems that often rely on manual configuration and hence are not practical.
As an example, the replacement of batteries of thousand of nodes could be extremely arduous or even impossible for structural health monitoring of civil infrastructures (i.e. bridges, towers). Hence, the solution to the growing burden of the system manager is automation, allowing the system to check its own status, to re-configure itself and fix the major problems in the network whenever it is possible. In this paper, we present and discuss the main features needed to achieve an autonomous large scale WSN. Furthermore, we compare these features with the state of the art of real-world large scale WSN deployments showing that further solutions are needed to drastically reduce human intervention while guaranteeing the main functionalities of the system.
\end{abstract}




\section{Introduction}
\label{sec:Introduction}
In the last decade, the wireless sensor network field has evolved to the point where it is possible to deploy sensor-node applications over a long period of time with the expectation that they will work to produce useful, scientifically-relevant data. Driven by the need to collect data about people behavior and health status, WSNs for health-care have emerged in recent years~\cite{health}. Similarly, WSNs has been applied to museums~\cite{museum} for preventive conservation of art purposes, to nuclear power plants for real-time monitoring~\cite{nuclear}, for precise agriculture~\cite{agriculture}, for structural health monitoring system of air-planes~\cite{airplane}, for car parking management system~\cite{car}, for oceanographic monitoring~\cite{ocean} and many others. The potential of WSNs has increased in the past few years with the advent of the Internet of Things (IoT)~\cite{IoT}. This broad term covers all non-computer and non-phone Internet-connected devices. Many in the business community view IoT devices as interesting consumer gadgets (i.e Fitbit, Apple Watch). And not surprisingly, IoT devices are coming into their own for business and industrial use, reinventing industries such as health-care and transportation~\cite{IoT2} as an example. In-fact, digitizing and streamlining the sharing of health or GPS data has the potential for dramatic gains in efficiency and save billions of dollars. 
Unfortunately, most of the work done in the past on WSNs is mostly based on simulation or on networks composed by a few numbers of sensor nodes. But in order to follow the current IoT trends, new techniques need to be implemented since orders of hundreds or thousands of sensor node devices are not manageable by today's WSNs systems that rely on manual configuration and hence are not practical. An exponential amount of time would be required to maintain the network, and introducing errors is almost inevitable. As an example, the replacement of batteries of thousand of nodes could be extremely arduous or even impossible for structural health monitoring of civil infrastructures (i.e. bridges, towers). Hence, the solution to the growing burden of the system manager is automation, allowing the system to check its own status, to re-configure itself and fix the major problems in the network whenever it is possible.
Similarly, scaling to thousands of nodes determines a growing number of failures that if not well managed can cause system overhead~\cite{fault3}.
Thus, the importance of WSNs in the mentioned application is fundamental but still, a lot of effort is needed to address problems such as the deployment of large-scale systems that meet the application's requirement even when operating in unsupervised environments~\cite{health}.
In fact, deploying these systems at large scale requires attention at both engineering and deployment side. Several recent attempts show successes and failures~\cite{failure1, failure2, failure3} offering importance guidelines on real-world deployments~\cite{success1, success2}.
In particular, while looking at real-world large scale WSN scenario, papers as~\cite{benny1,benny2, benny3} provide evidence on how to deploy a large scale network but no details are reported on how to maintain the network after the first deployment. This specific concern is evident in several white papers and case studies presented by different companies as Daintree Networks~\cite{Company1}, Cypress Envirosystems~\cite{Company2}, Millennial Net~\cite{Company3}, Echelon~\cite{Company4} and UTC technologies~\cite{Company5} that are leading companies in wireless solutions for smart building control. As an example, a case study conducted at the Salk Institute by Daintree Networks~\cite{Company6} shows that the installation of wireless real-time energy consumption controls allows to measure, manage and achieve up to 77\% in energy savings. But, no details are reported on how the company manages maintenance and failures in the network. 
This lack of details is not negligible since looking at the wireless occupancy sensor device~\cite{Company7} manufactured by the same company, authors report that a network join can be re-triggered manually at any time by pushing and holding a button. In the same tool, changing the PIR sensitivity requires a manual opening and closure of the cover. Hence, no automation is used to maintain the network but human intervention is always required.
Moreover, looking at several white papers from Echelon~\cite{Company8, Company9, Company10} authors report how important is the existence of guidelines to ensure that control devices share their information and are able to re-configure the network if needed. Hence, LON MARK Association~\cite{Company8} defines interoperability guidelines that would permit multiple technicians to simultaneously configure and maintain different portions of a control network using tools from different manufacturers. But again, no details are reported on how the company manage maintenance and failures to achieve a long-lived Wireless Sensor Network. In a case study by Echelon in~\cite{Company10} that recently completed a year-long investigation of RF-based technologies (ZigBee, Z-Wave, Millennial Net, and Dust), they found that the new RF technologies offered very poor robustness against sources of interference, very limited distance operation and mediocre battery performance. They tested them in real-world scenarios discovering that none of the radios could operate reliably at 30 meters. At distances as short as 10 meters, the radios had an insufficient operating margin to work reliably over time. Turning on noise sources dropped the operating distance so low that repeaters would be required every 5 to 8 meters. Hence, the company proposes their new system called Echelon Power Line, claiming that it is more than 99.7\% successful in messaging, range not affected by metal foil insulation, fault tolerance very high, extensive scalability and many others important features for a WSN. But again, no proofs are showed about its capabilities and evidence on maintenance for a long-lived system are not reported.
Finally, to confirm the importance of the mentioned problems, several attempts of large scale deployment of wireless sensor nodes have been made also inside houses~\cite{HomeOS} and studies of technology used in the home help explain the gap between the longstanding vision of connected homes and its reality ~\cite{HomeOS1, HomeOS2}.

This paper surveys prior work which has addressed on large scale deployment of WSN. The rest of the paper is organized as follows: Section \ref{sec:chapt2} provides an overall view of the features needed nowadays for achieving a long-lived autonomous large scale WSN.
Section \ref{sec:chapt3} describes literature contribution to solve the problems described in Section \ref{sec:chapt2} but focusing on real-world WSN. 
Section \ref{sec:chapt4} provides an overview of the system that has been implemented by our group and deployed on the second floor of the Computer Science Department at UC San Diego.
Finally, Section \ref{sec:chapt5} presents the challenges involved on large scale deployment of autonomous WSN and summarizes lessons learned from previous deployments and provides a brief discussion and future directions of research.

\section{How to: Long-lived Large-Scale WSN}
\label{sec:chapt2}
As previously mentioned, over the past years a lot of work has been addressed in the literature on improving WSN systems, but most of the projects have been evaluated only in small networks or within simulations. While looking at how to maintain a large scale WSN a lot of work is still missing.
In this section, we are going to present the main features needed to make WSN autonomous from human intervention. Furthermore, we are going to present how these features have been developed in the literature.
The success of autonomous large scale WSN is strongly related to the overcome of the following points:

\begin{itemize}
\item \textit{Energy Harvesting for Battery Lifetime:} the success of WSN and their ubiquitous use is constrained by energy supply which is generally provided by batteries. Thus, energy harvesting mechanisms must be taken into account to allow a long operational lifetime of WSNs and preventing the demanding work of replacing batteries that became an arduous challenge on large scale deployments of WSNs.

\item \textit{Aggregation:} the system has to be able to collect data from different sources, independently from the nature of the sensor measure and the communication protocol used.
\item \textit{Management:} the system has to be able to store, show and let the user manage the data in an easy manner.
\item \textit{Self-Repairing:} the system has to be able to manage the failure whenever it is possible and overcome the issues by itself through a self-repairing procedure.

\end{itemize}

These steps are fundamental in order to build a WSN able to continue to perform in a non-supervised manner. We now describe research efforts that attempt to solve the main features previously described: 

\subsection{Energy Harvesting}
One common problem to WSNs is the lack of reliable power for the remote sensor node in the network. Even in cases where power might be available, the cost and difficulty of wiring the WSN remote to the existing power mains could be prohibitive. Furthermore, many WSNs are installed some years after the original installation of the system to be monitored and even if power may be available, it could be too difficult to connect to the devices. For these reasons, most WSNs out in the field are (primary) battery powered. Unfortunately, capacity is an important limitation in using batteries to power WSNs since the current consumption demand of some WSN applications will require size and price of the batteries that are prohibitive to guarantee a long lifetime (years) of the sensor nodes \cite{harvesting10}.
Moreover, the use of big size batteries is not always allowed (house, building) and the use of smaller batteries would just decrease the time between their replacement. Furthermore, the replacement of batteries could not be always possible in specific applications like structural health monitoring of civil infrastructures, since the sensors could be installed in places extremely difficult to reach. The situation is even more complicated if the network that we consider is composed of thousands of sensors.
For these reasons, it is important to extract the energy from the environment, allowing a long operational lifetime of WSNs and preventing the demanding work of replacing batteries on large scale deployments. Thus, energy harvesting is becoming the solution for the development of autonomous WSNs \cite{harvesting1,harvesting2}. There are numerous energy harvesting technology available such as solar-panel, vibration harvester, thermopile, RF, photovoltaic and each of those has a specific harvesting capability as showed in Table~\ref{tab:Harvesting_mechanism}~\cite{harvesting1,harvesting10}.  

\begin{table}[ht]
\begin{center}
    \begin{tabular}{| c | c |}\hline
    Energy Source        & Harvested Power Levels \\ \hline
       \textbf{Vibration/Motion}     &             \\ \hline 
    Human                 &4$\mu$W/cm2    \\ \hline
    Industrial             &100$\mu$W/ cm2\\ \hline
    \textbf{Temperature Differential}&\\ \hline
    Human &25$\mu$W/ cm2\\ \hline
    Industrial& 1-10 mW/ cm2\\ \hline
    \textbf{Light}&\\ \hline
    Indoor& 10$\mu$W/cm2\\ \hline
    Outdoor& 10mW/cm2\\ \hline
    \textbf{RF Energy}&\\ \hline
    GSM &0.1$\mu$W/cm2\\ \hline
    Wi-Fi &1$\mu$W/cm2         \\ \hline                       
 \end{tabular}
\caption{Power gathered from different energy harvesting sources.}
\label{tab:Harvesting_mechanism}
\end{center}
\end{table}

Table~\ref{tab:Harvesting_mechanism} shows that the lack of energy generated is one of the key issues of energy harvested sources. For this reason, a critical topic of research in this area has been on power management. Efficient power management is important to maximize the benefits of having the extra harvested energy and a consequence of using these harvesting devices is the need of information about future energy availability that is absolutely required to perform optimal routing decisions. To achieve this, an environmental energy harvesting framework (EEHF)\cite{harvesting4} has been proposed. EEHF is a distributed framework for aligning the task distribution with energy availability.
\begin{figure}[ht]
    \centering
        \includegraphics[width=\linewidth]{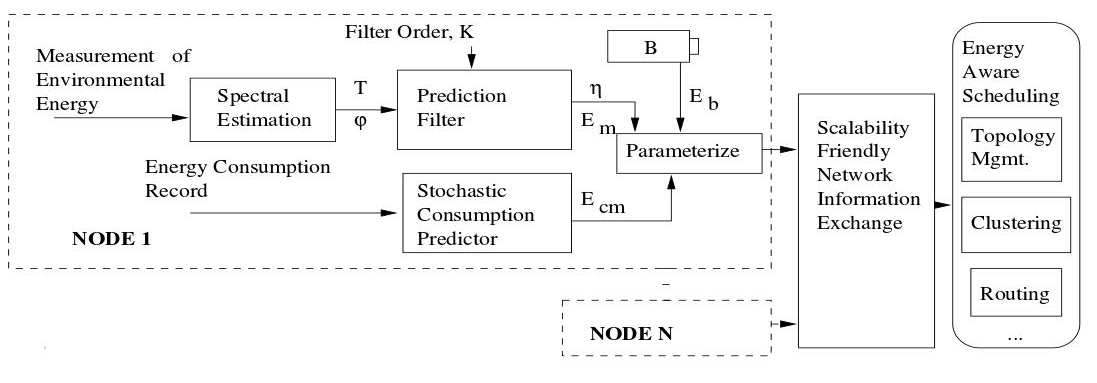}
    \caption{Interactions between various EEHF algorithms}
    \label{fig:energy}
\end{figure}
EEHF is composed by several blocks as showed in Figure \ref{fig:energy}:\\
\textit{(i) Spectral Estimation:} Learns the key spectral components in the availability waveform and generate the parameter \textit{T}, the time duration over which \textit{Em} (mean energy expected in subsequent T duration) is predicted. The waveform is then expected to repeat approximately every \textit{T} duration.\\ 
\textit{(ii) Prediction Filter:} Given \textit{T}, EEHF attempts to predict \textit{Em} for the duration \textit{T} in future.
A large variety of adaptive filters (least mean squares,  normalized, fixed coefficient) based on auto-regressive methods are available. This block also tracks the error in prediction and assigns a confidence value, $\eta$ to the prediction.\\
\textit{(iii) Stochastic Consumption Predictor:} This block tracks the average consumption, \textit{Ec}, in every \textit{T}.\\ 
\textit{(iv) Parameterize:} Combines the parameters learned by the above blocks, and the remaining battery \textit{Eb} into one cost metric to determine \textit{E}, the effective energy available.\\
\textit{(v) Scalability-friendly Information Exchange:} The previous blocks learned the temporal characteristics of the energy available at each node, locally. However, lifetime optimal scheduling requires global decisions to be made based on this information. Two approaches are proposed for sharing the local information which are both scalable with increasing network size and node density:\\
\textit{1. In-network Averaging:} 
Rather than explicitly transferring the effective energy available calculated \textit{E} at each node to a central scheduling entity, EEHF circulates the information about the average \textit{E} and maximum of \textit{E} among all nodes. The nodes can then volunteer to accept a workload L proportional to the energy availability consumption or go to sleep otherwise. In this way, the system avoids distributing the local \textit{E} at every node but only a few parameters making the network scalable.\\ 
\textit{2. Distributed Scheduler:} Certain tasking algorithms learn the local energy costs on their own as required. 
Then, a distributed route discovery algorithm chooses a minimum cost route without the framework having to explicitly provide for the sharing of information.\\
Authors compare the performance of EEHF and a residual energy based scheme, for routing.
Simulation results show that EEHF learn the energy environment and exploit better the energy resources improving the performance of the sensor network up to 200\% in lifetime.

In addition, to supplement the energy supply in battery-powered systems, energy-harvesting can enable a new mode of operation, namely, the energy neutral-mode \cite{harvesting5} in which the system uses only as much energy as it is available from the environment. In this mode, the power-management design considerations are very different from those of maximizing lifetime as for (EEHF) \cite{harvesting4}. In this case, the energy used is less than the energy harvested. The system has multiple distributed components each harvesting its own energy and the performance depends on the spatiotemporal profile of the available energy.
In particular, authors present and discuss an harvesting system in which the energy production profile is characterized as a
($\varphi$1, $\sigma$1, $\sigma$2) function, the load is characterized by a ($\varphi$2, $\sigma$3) function, and the energy buffer is characterized by parameters $\eta$ for storage efficiency, and $\varphi$ leak for leakage. The following conditions are sufficient for the system to achieve energy-neutrality: (i) $\varphi$2$\le\eta\varphi$1-$\varphi$leak; (ii) \textit{B0}$\ge\eta\sigma$2+$\sigma$3 and (iii) \textit{B}$\ge$\textit{B0}, where \textit{B} denotes the capacity of the energy buffer and \textit{B0} is the initial energy stored in the buffer.

Authors give also other important considerations for the design of an energy-neutral operational WSN:\\
\textit{Buffer Size and Related Considerations:} battery storage capacity degrades with multiple charge-discharge cycles \cite{harvesting6} and after 500 deep charge-discharge cycle the storage capacity of a NiMH battery falls to 80\% of its original. The use of a larger battery will slow down degradation significantly. Lions batteries use a high charging current that may never be supplied by the harvesting source while for NiCd batteries, the charging current is acceptable, but memory effect makes its use for partial charge and recharge cycles inappropriate. Ultra-capacitors have a high $\eta$, but also high leakage, which makes the charger much smaller than that achieved using batteries.

\textit{Achievable Performance Level:}
If the load consumes more power than the power produced by the energy harvesting source, its performance must be scaled down by using techniques such as duty-cycling among low-power modes or dynamic voltage scaling. If performance is battery dominated then power-management strategies need to be considered maximize lifetime.\\
\textit{The harvesting-aware power-management:} it consists of three parts: the first part is an instantiation of the energy generation model, which tracks past energy input profiles and uses them to predict future energy availability. The second part computes the optimal duty cycles based on the predicted energy.
The third part consists of a method to dynamically adapt the duty cycle in response to the observed energy generation profile in real time.

A similar work \cite{harvesting2} presents the energy constraints of traditional WSN by showing the power levels available from state-of-the-art energy harvesting devices that range from tens of $\mu$W to several mW (1\% to 20\% of operating power) which is not enough to power the sensor node continuously. Using the energy harvesting rates presented in \cite{harvesting7}, it is possible to estimate the duty cycle achievable by the Crossbow MICAz based on the power consumption requirements of 83.1mW in receive state and 76.2 mW in transmitting state. By considering the harvesting rate on a 10cm\textsuperscript{2} material which is about the same size as the mote, the estimation results are shown in Table \ref{tab:Harvesting}.

\begin{table}[ht]
\begin{center}
    \begin{tabular}{| c | c | c | c |}\hline
       & Power     & Energy &  Duty \\ 
      Technology   &     Density  &  Harvest    &  Cycle \\ 
                                            & (\small{$\mu$W/cm\textsuperscript{2}}) & Rate (mW)&  (\%)\\ \hline
    Vibration -      &        &      &           \\
    Electromagnetic        &    4.0    &   0.04   &          0.05        \\ \hline
    Vibration - &     &    &            \\ 
    Piezoelectric        &    500     &       5        &         6                \\ \hline
    Vibration -  &   &      &      \\ 
    Electrostatic        &    3.8     &       0.038        &         0.05            \\ \hline
    Thermoelectric                      &   60       &    0.6   &          0.72            \\ \hline
    Solar -   & &   &         \\ 
    Direct Sunlight      &   3700  &    37   &        45                \\ \hline
    Solar -   &   &     &   \\ 
    Indoor                    &   3.2  &    0.032       &         0.04            \\ \hline
    \end{tabular}
\caption{Achievable Duty Cycle by MICAZ with 10cm\textsuperscript{2} of harvesting material}
\label{tab:Harvesting}
\end{center}
\end{table}

\begin{figure}[ht]
    \centering
        \includegraphics[width=5cm]{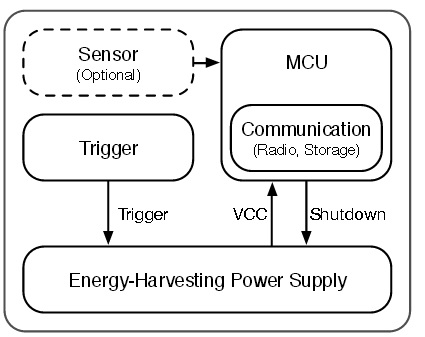}
    \caption{Energy-harvesting node architecture}
    \label{fig:campbell}
\end{figure}

Hence, current energy harvesting solutions are still in the development phase towards improvements in power output and are unable to provide sustained energy supply to enable WSNs continuously.
Energy harvesting for indoor environments could present further problems in the development of a large scale autonomous WSN. In fact, many WSNs for indoor environments are installed some years after the construction of the building (house, commercial buildings) creating further complication for scalability and installability \cite{harvesting9}. To address these issues, authors in\cite{harvesting9} propose an architecture to aid building-scale control systems in the future. In particular, Campbell et al, show that two are the main factors that limit how well sensors scale in buildings: 
\textit{Size:} the volume of the sensor node reserved for the power supply can be reduced to a battery-reserve able to gather energy at run-time and only storing enough for immediate or short-term operation.
\textit{Power:} the use of an intermittent source of power as energy-harvesting devices makes the design of building sensors capable of periodic sampling, event detection and other common features that can accomplish typical building monitoring tasks.
To achieve these goals the system has to be composed into the following four main subsystems: an energy-harvesting power supply (to gather and stores energy  to  power the sensor node), an activation trigger (responsible for initiating computation, and sampling a sensor), a sensing device (to capture some phenomena), and a data communication module (for transmitting the sensor data). The four subsystems are showed in Figure  \ref{fig:campbell}.

To verify and evaluate the energy-harvesting building monitoring architecture, authors implement three new sensors that represent three points in the design space: a vibration detector, an airflow meter, and a light/occupancy sensor. Each sensor is designed to monitor a particular phenomenon common to buildings and all of them are going to be powered by a solar harvesting power supply. Even if the three applications show an interesting accuracy, there are important drawbacks to consider: \textit{(i-Darkness)} photo-voltaic is not effective in dark conditions that occur when people leave the building or after the sunset. Thus, a rechargeable backing energy store as a battery may be useful for continuing to sense while the power supply is unable to harvest. But the use of battery increase size that is preferably avoidable for scaling in existing buildings. \textit{(ii Recharge Rate too Slow:)} the use of an intermittent power supply can limit the frequency of the system operation due to the recharge rate of the power supply. If the recharge rate is slower than the required sample rate, the sensor fails to operate correctly. Again, the use of a rechargeable battery can store the energy when there is harvesting surplus and use it when needed.
\textit{(iii Always On Receiver)}: sensors needs an always-on receiver in range for receiving transmissions to facilitate real-time sensing after collecting any data. In fact, the use of any complex MAC layer wireless protocols can decrease the performance of the system. To overcome this problem, authors propose the use of a long-range, sub-1 GHz radio that may reduce the number of receivers needed to cover a building or the use of the Bluetooth occupant's mobile phones as gateways for sending data to the cloud.

\subsection{Fault Detection and Self-Repairing}
Due to the use of large numbers of sensor nodes in WSNs, probability of sensor node failure gets increased, affecting the reliability and efficiency of WSNs. To maintain these important features, detection of failed or malfunctioning sensor node is essential. 
During the years, several researchers have proposed techniques for detecting failures in WSN deployments. \cite{fault} presents a method to detect the sensor node failure or malfunctioning by using the round trip delay (RTD) time to estimate the confidence factor \cite{fault5}of RTD path. RTD is the time required for a signal to travel from a specific source node through a path consisting of other nodes and back again. During the trip, if any sensor node fails or malfunctioning the time delays related to this sensor node will change and this introduces errors in estimating the round trip delay (RTD) times \cite{fault4}. Even if authors achieve good accuracy in detecting single malfunctions, the time required to check the network is increased by adding nodes and makes scalability arduous to reach. Furthermore, the power consumption of the nodes is increased by sending extra data and reduces the overall lifetime of the system.

A number of researchers have proposed proactive techniques to detect failures in WSNs in which each node reports its link status periodically as Memento \cite{fault6}. Memento collects the status of all the nodes in the sensor network in the form of bitmaps with the meaning of a particular health symptom. For example, in a status bitmap of type t, where t="alive", a bit pattern of 1101110 says that nodes 3 and 7 (the "0" bit positions) are not believed to be alive, while the others are. By the use of this semantics and by implementing a watchdog for that symptom, health monitoring modules control the bits in a local status. The Memento protocol calculates the aggregate result of each node by combining the node's local state. The protocol sweeps the entire network every $\tau sweep$, and delivers the global aggregate result to the gateway node that relays the information to the server which understands the semantics of each bitmap type. 
The proactive approach adds computational and communication costs at the nodes which significantly shorten network lifetime. Furthermore, it suffers from significant latency since status information is only sent periodically. To overcame these problems, researchers recently proposed the use of passive information collection for the purpose of failure detection \cite{fault7}. The passive approach consists of the idea that information useful for failure detection can be extracted from regular data packets sent to the sink node. By doing this, the passive approach does not incur significant network overhead but its accuracy depends on the sink node using non-deterministic means of inferring the operational status of nodes and links based on the data collected. As a result, the proposals suffer from poor accuracy and do not scale well with network size.

\begin{figure}[ht]
    \centering
        \includegraphics[width=7.4cm]{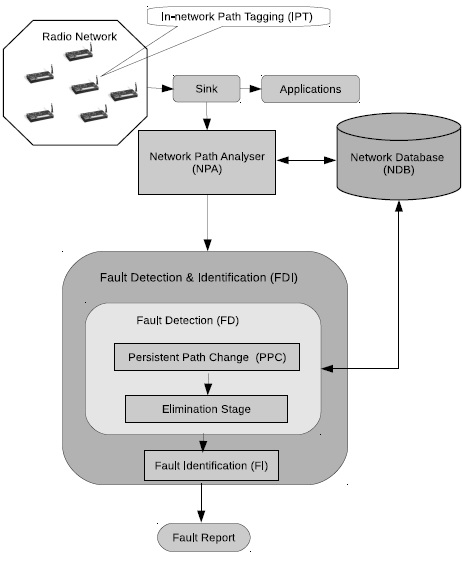}
    \caption{SBFD framework}
    \label{fig:SBFD}
\end{figure}

The combinations of a lightweight in-network packet tagging and server-side storage-intensive computation is presented in \cite{fault2} as SBFD (Sequence Based Failure Detection). The SBFD framework is depicted in Figure \ref{fig:SBFD}. The framework consists of four main components:\\
(i)In-network Packet Tagging (IPT),
(ii)Network DataBase (NDB),
(iii)Network Path Analysis (NPA),
(iv)Fault Detection $\&$ Identification(FDI).

\textit{Fault Detection $\&$ Identification (FDI):} IPT is performed by the nodes in the network. NPA and FDI are performed by, and the NDB is stored at the sink Node. During normal network operation, IPT causes a path checksum to be added to all packets sent from SNs(Source Node) to the Sink. Nodes on the routing path of the packet update the path checksum using their node ID and the current path checksum as inputs to the Fletcher checksum algorithm. When a packet arrivals at the sink, the path checksum is used to determine the packet path by means of look-up in the NDB. The NDB is pre-populated with the paths and path checksums for the network based on its known topology and NPA. Then, the FDI module inspects the NDB to determine if the network path statistics have changed and if they have, the FDI module reviews the NDB path information to determine
if a fault is likely. If a fault is suspected, the FDI module sends control messages to the
affected nodes. Based on the responses to these control messages, the Sink determines
if a fault has occurred or not. If a fault is detected the location and type of the fault are
reported. The types of faults detected are Node Failure, Link Failure, and Node Reboot.

Even if the SBFD is lightweight in terms of communication overhead and it is efficient in terms of node computation, it cannot catch the presence of dumb nodes in the network and hence no algorithm is implemented in order to recover the data once the connectivity has been re-established. Furthermore, even if the SBFD is scalable and allows to detect failures at multiple levels, it is missing a clear approach on how to sustain a network after a fault is detected that is a priority for large scale autonomous WSNs. Self-repairing techniques are needed to correct the faults once detected allowing the system to check its own status and to re-configure by fixing the problem in the network. To do that, the system has to be able to check for updates that can solve problems whenever possible. In subsection \ref{subsec:piloteur} I am going to verify a recent technique that allows self-repairing in WSN.\\

\subsection{Aggregation and Management}
\label{sec:aggregation}
Once the data have been generated by the sensors, an organized structure to manage the data collected and provide the information required to the different control subsystems (i.e. HVAC and lighting in buildings) is needed. The importance of such a structure for a large scale sensor network is fundamental because thousands of sensors that continuously produce relevant data can be extremely hard to manage if the system does not scale well. Furthermore, when thousands of sensors participate in a network, different sensors nodes equipped with different communication protocols can be present. A clear organization of the data can also facilitate monitoring and fault detection techniques that check for the correctness of the data and hence enabling maintenance. Hence, a scalable system that can aggregate and manage information from different sources is absolutely required while developing a large scale autonomous WSN.
To overcame these problems, different solutions have been presented in the literature and I am going to present the most relevant ones: Building Depot\cite{BuildingDepot}, HomeOS~\cite{HomeOS}, and Sensor Andrew project\cite{SensorAndrew}.

\textbf{HomeOS:} 
Authors in~\cite{HomeOS} have developed a PC-like abstraction for technology in the home. All devices in the home appear as peripherals connected to a single logical PC. The HomeOS addresses three objectives: \textit{(i)} it provides a central management platform for all the devices for a
non-expert user, \textit{(ii)} it provides a platform for application development which can be easily configured in different kinds of home environments, devices, and user control. \textit{(iii)} it makes easy to add a new device to the system. Their current prototype supports several device protocols (e.g., Z-Wave and DLNA) and many kinds of devices (e.g. lights, media renderers, and door/window sensors).
Thus, HomeOS is a perfect example of how to integrate different types of sensors and provide an interface to the occupant that is easy to use. However, the framework is meant for homes, and scaling issues that come up for large scales (i.e. buildings) are not addressed. For example, data collected by the system can be used by a user to develop applications that control different appliances. Hence, the system is not meant to control and manage orders of thousands of devices in an autonomous way but requires human maintenance.

\textbf{Sensor Andrew:} 
The Sensor Andrew~\cite{SensorAndrew} system addresses problems similar to HomeOS but in a commercial building environment such at a university-wide scale. The architecture is built around the Extensible Messaging and Presence Protocol (XMPP). It is a standard scalable messaging and presence protocol with user/group authorization, authentication and access control. The system utilized a Transducer Layer that provides adapters to Internet-
connected devices from all the end-point sensors and actuators. Furthermore, it is responsible for supporting all the different types of devices and communication protocols.
The Gateway Layer consists of devices which have access to the Internet and are configured as XMPP clients. The gateway collects all the information from the transducer layer using low-level protocols and passes it on the Server Layer using XMPP through event nodes. Devices in the server layer need to subscribe to event nodes to receive published data.
The system uses a web application, called Data Handler, to maintain the schema, business rule and read/write functions. It provides the
interface for browsing, editing, creating transducer and device metadata records in the registry.
Rowe et al. demonstrate the capability of the Sensor Andrew system in a home environment where energy meters and motion sensors are deployed throughout a home and are used to infer the appliances which show a high correlation between motion and energy use. 
The Sensor Andrew system provides a good communication framework among sensors deployed in a large scale deployment. However, an evaluation of the system in a real deployment has not been shown in the literature since the applications show-cased were developed in a home setting. 
There is no structure given to the data obtained from different sources, for querying status of sensors, ease of development and maintenance.

\textbf{Building Depot:}
Building Depot \cite{BuildingDepot} was developed for commercial building and allows storage, access, and sharing of data from sensor nodes. Building Depot uses RESTful HTTP service, similar to the ideology behind sMAP \cite{smap}, and uses JSON objects for data representation.
The data packet sent by the sensor nodes to the Base Stations are collected by Building Depot through the \textit{Data Connectors} that are developed for each of the protocols used. In this way, sensors that are using the same protocol can reuse the same connectors. Once collected the data are organized by the following structure:\\
\textit{Data Service} is the module which handles all the sensor data and makes it available to different applications. The module also stores contextual data for each sensor(i.e. location, type of data,)\\
\textit{Directory Service} contains links to the various Data Services and the Directory Services beneath it. These Directory Services are used to form a hierarchical tree of all of the Data Services and Directory Services that make up an institutions Building Depot system.\\
\textit{User Service} provides user management to ease data sharing and administration. Users are authenticated using a registration system similar to most personalized systems on the internet today. Hence, Building Depot provides a management framework to store the vast amount of data collected from different sources in an organized manner and provides a uniform interface to access this data. The drawback of the Building Depot system is that it has not been evaluated in a realistic environment.

\begin{table}[ht]
\begin{center}
    \begin{tabular}{| c | c | c | c |}\hline
    \small{Features/System }    & \small{HomeOS}& \small{SensAndrew}    &  \small{BuildingDepot}    \\ \hline
    \small{Aggregation}            &    x        &     x               &          x    \\ \hline
    \small{Management}            &    x        &                   &         x    \\ \hline
    \small{Scaling    }            &            &       x            &         x     \\ \hline
    \small{Maintenance}            &            &                    &         x     \\ \hline
    \end{tabular}
\caption{Building Depot, SensorAndrew and HomeOS for solving Aggregation, Management, Scalability and Maintenance problems in LS autonomous WSN.}
\label{tab:table_Comp}
\end{center}
\end{table}

\section{Real-World Autonomous Large-Scale WSN}
\label{sec:chapt3}

In this section, we present the state of the art of real-world large scale deployments of wireless sensor networks. In particular, we are going to focus on how the following WSNs manage the features presented in Section \ref{sec:chapt2} in order to become autonomous.

\subsection{Cross-rails:}
\begin{figure}[ht]
    \centering
        \includegraphics[width=\linewidth]{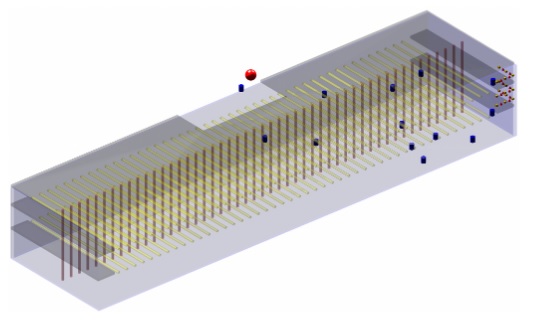}
    \caption{3D model of Paddington station main box and WSN layout.}
    \label{fig:crossrail}
\end{figure}

Authors in \cite{crossrail} deploy a WSN in an excavation for a new Crossrail station at Paddington, London. This excavation takes the form of an underground box that is 260m long, 25m wide and 23m deep. 
The main aim of the WSN deployment was to monitor deformation of three diaphragm wall panels on one of the corners of this underground box during excavation as showed in Figure \ref{fig:crossrail}. Wireless tilt and displacement sensors were installed to measure inclination, angular distortion and relative displacement of these corner panels at two different depths. These measurements can potentially offer some insights on the real performance of a box corner during large deep excavation. The WSN gateway was positioned outside the underground box, as it requires a power supply and good 3G signal coverage. Several relay nodes were also attached to the diaphragm wall panels and plunge columns.

\begin{figure}[ht]
    \centering
        \includegraphics[width=\linewidth]{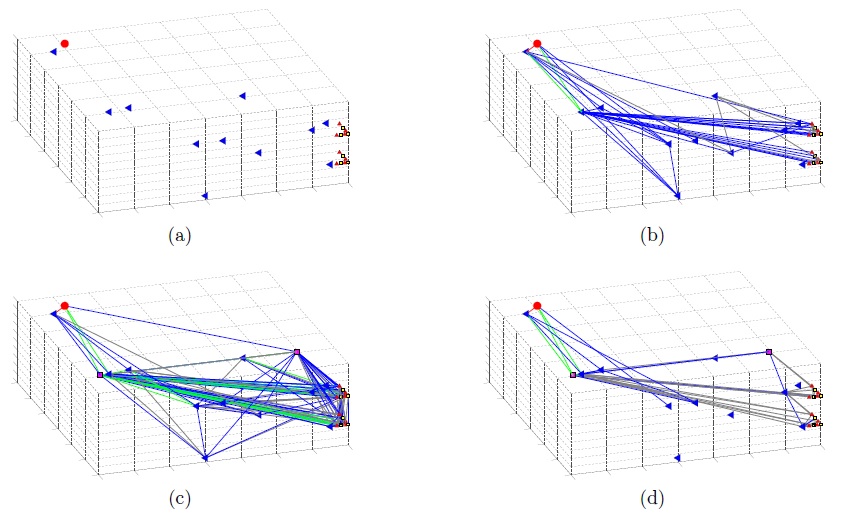}
    \caption{Network layout: (a) Initial network layout; (b) Initial topology; (c) Two relay nodes were added to introduce path diversity; (d) Final network topology at the end of six months. Link color represents the average number of connections made to the gateway per day during the 5-day period. Grey line indicates one-two connections; blue line, between two and 20 connections; green line, between 20 and 200 connections; and red line, more than 200 connections.}
    \label{fig:dynamics}
\end{figure}

\textbf{Network}: The tilt and displacement sensors used in the deployment were obtained from Wisen Innovation. These sensors are commercially available and designed for use on construction sites and are packaged in robust metal housings. Internally the devices are based on the AVR ATmega1281 processor and the IEEE 802.15.4-compliant AT86RF231 radio. Fifteen sensors measured displacement while twelve equipped with inclinometers. Thirteen relays were used in total.
The gateway used a Memsic Iris mote acting as the root node and border router. The use of the gateway board allowed for resetting or reprogramming the Iris remotely if required.
The application software running on the wireless sensor devices was developed in Contiki OS [1]. Nodes use Contiki's standards-based IPv6 stack (6LoWPAN/RPL) for link-local addressing and routing, and ContikiMAC atMAC layer for low-power operation.
Each node was initially programmed to send link-local information periodically to assist in identifying and diagnosing potential network failures. 
The layout of the wireless sensor network at the Paddington excavation is shown in Figure \ref{fig:dynamics}b). 
Sensor data messages are sent from each node at fifteen minutes intervals. The network experienced continuous connectivity problems that resulted in data message delivery ratios of below 10\% in the two first months after deployment. Even if diagnostic was limited to analysis of those messages that successfully reached the gateway, it did provide some insight into the actual network behavior. Figure \ref{fig:dynamics}b) shows the network topology constructed from the diagnostic messages sent by all nodes in the WSN. Interestingly, it shows that sensor nodes were routing messages via a single far relay which located on the opposite side of
the station box in close proximity to the gateway, rather than using nearby relay nodes to forward messages. 
A potential reason for this was thought to be the Wall Antenna effect which would have limited radio propagation in this direction. This prompted the installation of two additional relay nodes in an attempt to increase path diversity and reduce the routing overhead at the far-off relay node.
Figure \ref{fig:dynamics}c) shows the new network topology after the addition of relay nodes shown as magenta squares. 
With the installation of the two additional relays, an improvement in data message delivery ratio for all sensor nodes
(up to three times more) was observed. Unfortunately, this improvement only lasted for around twenty days, after which
the message delivery ratios dropped again. Figure \ref{fig:dynamics}d) shows the network topology constructed from the limited amount of diagnostic data reaching the gateway.\\ 
In a network suffering from high message loss, sending diagnostic information to the gateway does not provide any detailed insight into the cause of these losses. Therefore, all the nodes in the network were reprogrammed with new application software. This new application stores diagnostic data in local non-volatile (flash) memory rather than sending this information via lossy links. This stored data may be retrieved during repeated site visits.

Hence, despite the development of the core communications protocols necessary to enable wireless sensor networking, deployments on real-world sites can still be problematic. In the absence of suitable diagnostic data available at the gateway
node, it can be difficult on a large site to quickly diagnose and fix issues such as poor point-to-point link quality caused
by sub-optimal relay placement. Support tools to assist with the rapid deployment, diagnosis, and maintenance of such WSNs are required.

\subsection{Lab of Things:}
Authors in \cite{labthings} developed a shared infrastructure for home environments, called Lab of Things with the goal of lower the barrier of developing and evaluating new technologies for the home environment. Lab of Things (LoT) is a flexible platform that uses connected devices in homes. It provides a common framework to write applications and has a set of capabilities beneficial to field deployments including logging application data from houses in cloud storage, remote monitoring of system health, and remote updating of applications if needed (e.g. to change to a new phase of the study by enabling new software, or to fix bugs). 
Each household in Lab of Things runs HomeOS on a dedicated computer, the Home Hub, which interacts with the in-home sensors and hosts the applications needed for the study in which the household is participating.

\begin{figure}[ht]
    \centering
        \includegraphics[width=0.8\linewidth]{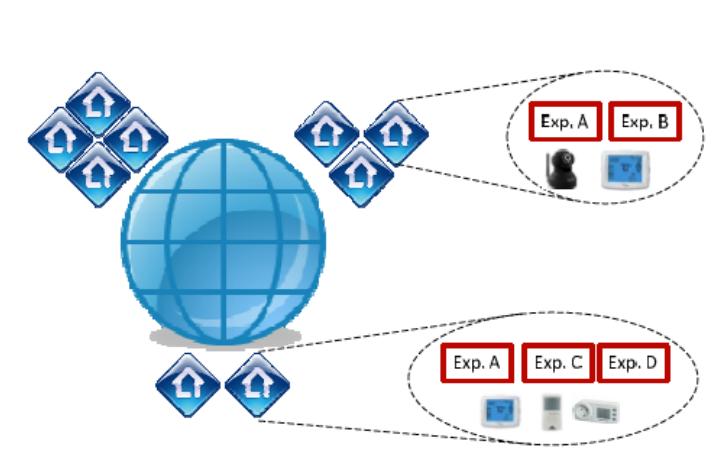}
    \caption{Lab of Things will consist of multiple sites across the world. Each site has multiple homes and is managed by a research group.}
    \label{fig:LoT}
\end{figure}

Overall, LoT enables:
\begin{itemize}
\item \textit{Extensible support for devices:} a wide range of devices (e.g., many Zwave sensors, IP cameras) can be easily interconnected by researchers to implement application scenarios.
\item \textit{Scaling:} field studies at scale through cloud services that can monitor and update experiments, and provide easy access to collected data.
\item \textit{Data storage and access:} Lab of Things incorporates a Home Data Store which provides a seamless transfer of application data to the cloud and handles scarce in-home storage and intermittent network connectivity. 
\item \textit{User security and privacy:} Studies involving end-user applications can easily incorporate robust user authentication and authorization.
\end{itemize}


\subsection{Piloteur}
\label{subsec:piloteur}
In \cite{Piloteur}, authors present the design and implementation of Piloteur: a lightweight platform for robust pilot studies of sensing and control systems in the home. Upon installation, it provides end-to-end monitoring of system operation, it repairs problems whenever possible, and it alerts the operator of fatal problems that cannot be repaired. In the paper, authors present the design and implementation of Piloteur as well as their experiences using it to deploy over 180 endpoints across 45 different homes. Despite thousands of potentially fatal node failures, Piloteur automatically repaired nodes most of the time reducing the physical human intervention.\\
\textit{Piloteur Under The Hood:} To provide the simpler interface, four separate Piloteur subsystems must operate together in a coordinated manner. The four subsystems are illustrated in Figure~\ref{fig:piloteur}:

\begin{figure}[ht]
    \centering
        \includegraphics[width=0.8\linewidth]{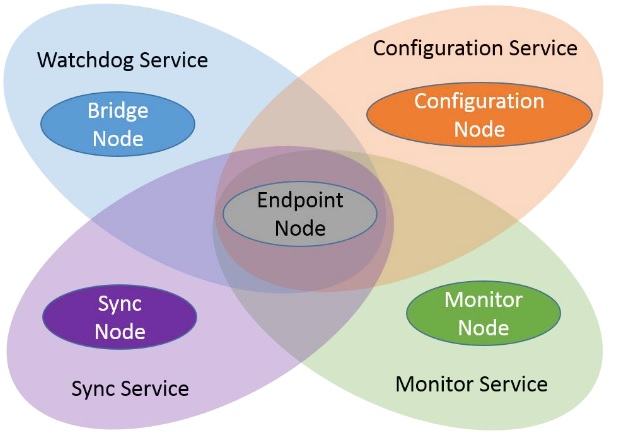}
    \caption{Piloteur architecture}
    \label{fig:piloteur}
\end{figure}

\begin{itemize}
\item \textit{The configuration service} is the first to be activated: it finds a nodes configuration file on the configuration node, downloads the file, and configures the endpoint and its drivers. Then, it continuously checks for configuration or software updates. The configuration service helps scale deployments up to dozens or hundreds of nodes with only marginal increases in setup and configuration time, and it can run on low-cost embedded devices such as the Raspberry PI for cost-effective scaling.
\item \textit{The watchdog service} continuously monitors both the hardware and software of the endpoint, including hardware drivers, peripheral devices, network connections, and Piloteur subsystems. It also maintains a reverse SSH tunnel to the bridge node, in case manual intervention is required. It records all status information in files in a logs directory and, if a problem is detected, corrective action is taken if possible. 
\item \textit{The sync service} continuously mirrors files from the data directory and logs directory onto the sync node. It then automatically purges old data in order to maintain sufficient local storage for new data. 
\item \textit{The monitoring service} continuously monitors both the data files and operational logs from each endpoint, provides a RESTful Web interface that indicates the status of the endpoints, and sends alert messages to the user if a critical problem is detected. It does not run directly on the endpoint.
\end{itemize}

Authors performed the efficacy of Piloteur in two pilot studies by operating for approximately 4 months. 
They demonstrate the efficacy of the Piloteur design with a case study in which deployed over 180 endpoints across 45 homes. The analysis indicates that Piloteur reduced maintenance visits to about 35\% of the actual number of failing endpoints. Despite thousands of potentially fatal software failures, only 26 physical maintenance visits were required to keep over 75\% of the nodes operational.


\subsection{Berkeley Deployment (@Scale)}
Dawson-Haggerty et al. \cite{@scale} present a year-long study, with 455-meters deployment of wireless plug-load electric meters across 4 floors in a large commercial building at Berkeley. For the study purpose, the authors implemented a staged stratified sampling method to analyze Miscellaneous Electric Loads (MELs) in the building. To provide stability to the wireless network, Load Balancing Routers (LBRs) were installed at strategic locations to act as routers for the IPv6 enabled energy meters. The LBRs were also used to facilitate remote debugging. They had overlapping regions to increase reliability in case an LBR breaks down. The authors use IPv6 to develop a compact implementation of services and to scale to large numbers.
They also take to a new level inexpensive energy metering technology similar to that used in the past. The science goals of the deployment require consideration of the accuracy required by the meters; they developed an extensive automated calibration procedure that could be efficiently applied to hundreds of meters to achieve better than 2\% accuracy.
The software on the meters was designed such that simple configuration parameters like sampling rate and calibration parameters can be extracted and changed on the fly. The authors found this utility extremely useful and had to seldom resort to re-programming the devices using over-the-air image update. Hence, their key insight was the value of configuration, not reprogramming. 

\section{The Occupancy Detection \\Network at UCSD}
\label{sec:chapt4}

\begin{figure}[ht]
    \centering
        \includegraphics[width=0.8\linewidth]{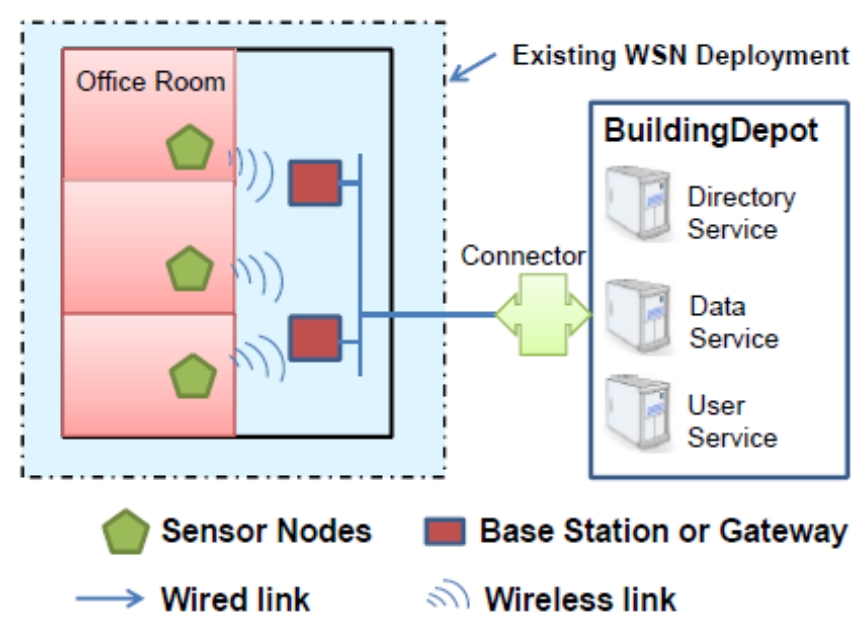}
    \caption{Building Blocks of the UCSD Network\cite{BuildingDepot}}
    \label{fig:Network}
\end{figure}

In this section, we describe our architecture and the monitoring framework used to obtain our results. Overall, our network is composed of three main parts: a basic sensor-node, the base station and the monitoring-storage system as shown in Figure \ref{fig:Network}.

\subsection{Building Occupancy System}
The primary goal of the Wireless Sensor Network deployed at the second floor of the Computer Science Department at UCSD is to detect occupancy inside the building. By doing this, the network acquires the necessary information to turn on the HVAC system in a specific area only when it is needed allowing energy saving in the building.\\
The design of this occupancy detection system\cite{UCSD} is based around several key objectives. First, authors
wanted to make the system as low cost as possible to decrease the expense deployment across a building-wide scale. Second, the system has to be incrementally deployable within existing buildings, without requiring large scale modifications such as new wiring. Finally, the occupancy detection algorithms should be very accurate in minimizing false-positives (which increases energy use), and more importantly, false-negatives (which may lead to discomfort) when controlling the HVAC system.

\begin{figure}[h]
    \centering
        \includegraphics[width=5cm]{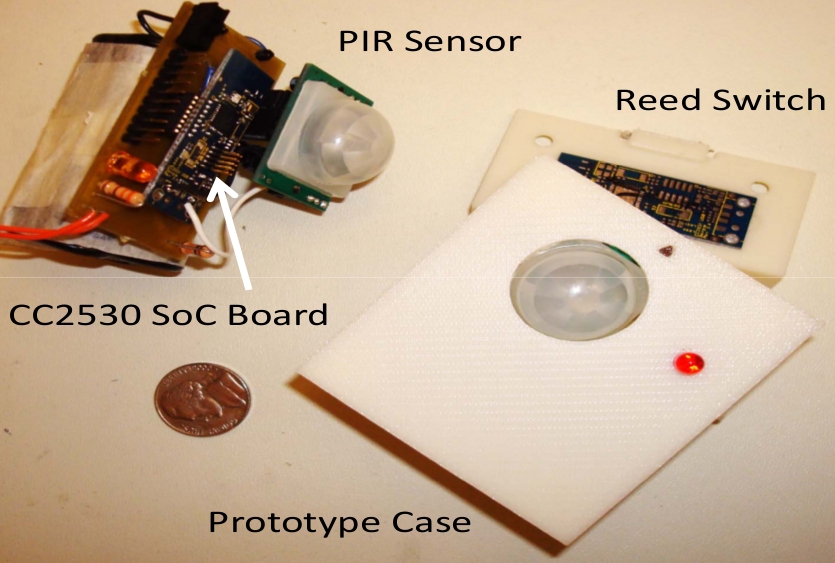}
    \caption{Synergy Occupancy Sensor presented in \cite{UCSD}}
    \label{fig:end_device}
\end{figure}

\textbf{Synergy Presence Node Design:} to satisfy the previous requirements the occupancy sensor node uses a combination of a magnetic reed switch door sensor and a PIR sensor module that enable highly accurate occupancy detection. For deployability reasons, the natural choice was to use a wireless solution that uses a TI CC2530 System-on-Chip integrated with a 8051 microcontroller core with and an 802.15.4 standard compatible radio in a small footprint and low-cost package. 
The board, batteries and the sensors are all placed in a case and Figure \ref{fig:end_device} shows a photo of the prototype. The overall cost of each board is under 15\$.

\textbf{Occupancy Detection Algorithm:} The combination of the reed switch and the PIR sensors together improve the accuracy
of the occupancy detection. The reed switch is able to sense when the door is open or closed. When the door is open, we
mark the room as occupied since for the authors in typical office building an open door denotes the occupant being in the office or being somewhere nearby. Similarly, when people leave an extended period of time (such as the end of the day) they typically close the door to their offices for security reasons.
When a door close event happens, there are two possibilities.
Either the person closed the door and headed out (room
unoccupied) or the person just closed the door and is still
inside the room (room occupied). To disambiguate between
these two cases we use the PIR sensor to determine if someone
walked near the door. If the PIR sensor goes high this
means that there is still a person in the room and we mark
the room as occupied. If the PIR sensor does not detect motion,
then we decide that there is no occupant in the room.
One limitation of this algorithm is that if a visitor closes the door while the main occupant of a room is sitting still at his desk (such as typing on the computer), the PIR sensor will likely not detect movement and thus determine the room to be
empty. Hence, once the door is closed the PIR sensor remains active and the capture of a movement will be labeled as an occupancy event.

\textbf{Wireless Network and Occupancy Server:} the wireless network infrastructure has been designed as a tiered topology.
Each of the wireless Synergy presence nodes communicates with their local base station over the 802.15.4 radio using the ZigBee protocol(free of cost and is certified by the ZigBee Alliance). The network is self-organizing. When a Synergy presence node powers on, it automatically connects to the nearest base station and will send occupancy data to its parent base station.
There can be multiple base stations on a floor of a large building, and each base stations is connected to the building
server network using either Wi-Fi or Ethernet. The occupancy data can then be analyzed in real time to determine the appropriate actuator actions(i.e run the HVAC or reduce cooling during unoccupied periods).
Finally, authors have implemented a website that allows this occupancy information to be observed in real time.

\subsection{The Base Station or Gateway}
Once the data are generated, the main function of the Base Station is to collect the data from the sensor nodes around the building and to make them available to the storage system. In our particular architecture, the Base Station is composed of a Raspberry PI. Each Base Station mounts four coordinators which are responsible to communicate with the sensor nodes and to print the received wireless packets to the Raspberry PI through the UART. Thus, every time a coordinator receive data from a sensor node, this information will be available in the memory of the Raspberry PI and hence it is ready to be gathered by the storage system.

\subsection{Building Depot}
Building Depot \cite{BuildingDepot} is the last building block of the system and allows storage, access, and sharing of data from the sensor nodes. It has been already described in section \ref{sec:aggregation}

Overall we can claim that Building Depot provides a management framework to store the vast amount of data collected from different sources in an organized manner, and provides a uniform interface to access this data.

\section{Conclusion and Future\\ Challenges}
\label{sec:chapt5}
In this paper, we analyzed the main features needed to achieve an autonomous large scale WSN. Furthermore, we compare these features with the state of the art of real-world large scale deployments analyzed in Section \ref{sec:chapt3}. The result of the analysis is showed on Table~\ref{tab:Hetero}.

\begin{table*}[ht]
\begin{center}
    \begin{tabular}{| c | c | c | c | c | c |}\hline
    Features/Systems                &Cross-Rail    & @scale     & Piloteur     &  Lab of Things  & OurSystem \\ \hline
    Aggregation                         &&            &       x        &         x        &     x        \\ \hline
    Simple setup                     &&          &                    &         x        &             \\ \hline
    Self-repairing                  &&           &    x       &                   &             \\ \hline
    Management                      &&   x     &    x           &        x        &    x        \\ \hline
    Scalability                      &&   x      &    x      &         x        &    x        \\ \hline
    Remote access for fixes and updates&x&     x  &      x        &    x        &    x        \\ \hline
    Simple Application Development    &&           &              &         x        &             \\ \hline
    Energy Harvesting                &&           &              &                 &             \\ \hline

    \end{tabular}
\caption{Features in real large-scale deployment WSN}
\label{tab:Hetero}
\end{center}
\end{table*}

Table~\ref{tab:Hetero} shows that all the real-world large scale network deployments analyzed in this paper do not meet all the requirements needed for an autonomous WSNs, and hence periodic maintenance and human intervention are always required for perpetual operations of the networks. In particular:
\begin{enumerate}

\item\textit{ Energy-Harvesting:} commercial energy harvesting available systems are cumbersome, expensive and still in the development phase towards improvements in power output that in the of the art presents a few tens of $\mu$W. Hence, the current state of technology in energy harvesting is still unable to provide sustained energy supply to enable WSNs continuously. For these reasons, no one of the real world WSNs that we analyzed exploited energy harvesting solutions.

\item \textit{Self-Repairing:} after a failure has been detected at any level, the system has to be able to manage it and recover the main functionalities whenever it is possible. In the past years, numerous fault detection techniques have been developed in the literature but self-repairing methods need to be improved in order for a network to long-live and reach autonomy. Piloteur is the only system that we found on a real-world deployment that was able to partially solve the mentioned problem. Even if only 35\% of failures were fixed without human intervention, we believe that this is a good starting point in the design of self-repairing WSNs.


\item \textit{Network Scaling, Aggregation and Management:} The management of thousands of sensors from different sources that continuously produce relevant data can be extremely hard if the network and system do not scale well. A clear organization of the data can also facilitate monitoring and fault detection techniques that check for the correctness of the data and hence enabling maintenance. Thus, a scalable system that can aggregate and manage information from different sources is absolutely required while developing a large scale autonomous WSN. Three out of the four real-world WSNs analyzed are able to scale to a large number of devices and manage information about thousands of sensor nodes. Furthermore, these networks allow the system to include data coming from different sources using different protocols. 









\end{enumerate}





\end{document}